\documentclass[
    ,final            
  ]
  {aipproc}

\layoutstyle{6x9}

\newcommand{\be}{\begin{eqnarray}}
\newcommand{\ee}{\end{eqnarray}}

\def\simge{\mathrel{%
   \rlap{\raise 0.511ex \hbox{$>$}}{\lower 0.511ex \hbox{$\sim$}}}}
\def\simle{\mathrel{
   \rlap{\raise 0.511ex \hbox{$<$}}{\lower 0.511ex \hbox{$\sim$}}}}
\newcommand{\kk}{k_\perp}

\begin{document}

\title{Cronin effect and high--$p_\perp$ suppression from
the \\\vspace{0.2cm}
Color Glass Condensate\footnote{Invited talk, presented at the
International Workshop IX Hadron Physics and VII Relativistic Aspects of
Nuclear Physics (HADRON-RANP 2004), Angra dos Reis, Brasil, 
March 28--April 03, 2004}}

\author{Edmond Iancu}{
  address={Service de Physique Th\'eorique, CEA/DSM/SPhT,
91191 Gif-sur-Yvette Cedex, France}
}

\begin{abstract}
I give a pedagogical survey of the nuclear collective effects
associated with gluon saturation and their impact on particle
production in high--energy proton (or deuteron)--nucleus collisions
at RHIC. At central rapidity, the theory 
predicts a Cronin peak due to independent multiple scattering off
the valence quarks in the nucleus. At forward rapidities, 
the peak flattens out and disappears very fast, because of the 
correlations induced through quantum evolution
in the nuclear gluon distribution at small $x$. Also, the ratio 
${R}_{\rm pA}$ between the particle yield in proton--nucleus and
proton--proton collisions is rapidly suppressed when increasing
the rapidity, because of saturation effects which slow down the
evolution of the nucleus compared to that of the proton. This
behaviour could be responsible for a similar trend observed in
the deuteron--nucleus collisions at RHIC.

\end{abstract}

\maketitle


\section{I. Introduction}

When accelerated up to RHIC energies ($\sim 100$ GeV/nucleon), a heavy
ion such as a gold nucleus is expected to `evolve' into a high--density
form of hadronic matter --- the ``Color Glass Condensate''  (CGC)
\cite{MV,CGC,CGCreviews} --- whose properties are qualitatively different
from those of `ordinary' partonic systems at lower energies.
This is the matter made of the small--$x$, `saturated',
gluons, and is characterized by an intrinsic
momentum scale, the {\it saturation momentum} $Q_s$ \cite{GLR},
which rises rapidly with the energy and the atomic number $A$ 
($Q_s^2(x,A) \sim x^{-\lambda} A^{\delta}$ with $\lambda\sim 0.3$
and $\delta\sim 1/3$) and acts as an infrared cutoff in momentum integrals
involving the gluonic spectrum (see \cite{CGCreviews} for recent reviews).
As usual, the variable $x$ denotes the longitudinal momentum fraction
of the `interesting' gluons --- those which participate in the
scattering ---, and decreases rapidly when increasing the energy
of the collision. Thus, for sufficiently small $x$ and/or large
$A$, $Q_s(x,A)$ is a {\it hard} scale ($Q_s^2\gg\Lambda^2_{\rm QCD}$),
which allows us to rely on perturbative techniques to get insight into
the properties of the CGC and of the high energy scattering in QCD.

Of course, the CGC remains `virtual' (it is a part of the nuclear
wavefunction) before any scattering takes place, but one expects
it to be `liberated' in a high--energy nucleus--nucleus 
(or proton--nucleus) collision,
and thus to determine the properties of the partonic system which
is created immediately after the collision \cite{CGCreviews,GM04}.
This system must be strongly interacting (because of its high energy
density and the many intervening degrees of freedom),
and also highly off--equilibrium (in particular,
because of the strong anisotropy in the initial distribution of
energy and momentum), so it is expected to undergo a violent
evolution, characterized in the early stages by a rapid longitudinal
expansion and a redistribution of the energy,
momentum and other quantum number among the various modes,
possibly leading to a thermalized state
(the ``Quark--Gluon Plasma'') at intermediate stages \cite{GM04},
which then further expands and cools down, until
it hadronizes into a multitude 
of hadrons (a few thousands) that are  eventually
captured by the detectors at RHIC \cite{QM02,QM04}. It appears therefore as
a challenge for the theorists to imagine observables which could
survive (almost) unchanged to such a violent evolution,
and thus carry out information about the initial conditions
(in particular, about the color glass).
It is furthermore a challenge for the experimentalists to
extract and measure such observables with a significant accuracy.

One of the observables which look most promising
for a study of the initial conditions (since less sensitive to
 `final state interactions') is the total multiplicity of produced particles
\be\label{dNdy}
\frac{{\rm d}N}{{\rm d}\eta}\,=\,\int {\rm d}^2p_\perp\,\,
\frac{{\rm d}N}{{\rm d}^2p_\perp {\rm d}\eta}\,,\ee
where ${\rm d}N/{{\rm d}^2p_\perp{\rm d}\eta}$ denotes the
spectrum of the produced hadrons (of a given species, or summed
over several species), $p_\perp$ is the hadron transverse 
momentum, and $\eta$ is its (pseudo)rapidity :
\be
\eta\,\equiv\,\frac{1}{2}\,\ln\,\frac{p+p_z}{p-p_z}
\,=\,\frac{1}{2}\,\ln\,\frac{1+\cos\theta}{1-\cos\theta}
\,=\,-\,\ln\,\tan\,\frac{\theta}{2}\,,\ee
with $p=\sqrt{p_\perp^2+p_z^2}$ and $\theta$ the angle
between the direction of the produced hadron and the collision
axis. It is an experimental fact that the integral in
Eq.~(\ref{dNdy}) is dominated by small momenta. In fact,
in lowest order perturbation theory, the spectrum
diverges at small $p_\perp$ like $1/p_\perp^4$, and the
corresponding integral is ill defined. But the non--linear effects
encoded in the CGC --- which, from the point of view of perturbation
theory, correspond to all--orders resummations of
high parton density effects --- provide an
infrared cutoff in the form of the saturation momentum $Q_s$,
which at RHIC is estimated as $Q_s\sim$ 1 GeV. Note that this
is marginally a hard scale, which justifies a posteriori the use
of perturbation theory in the construction of the theory for the CGC
\cite{CGCreviews}.

One sees that the very calculability of the total
multiplicity at RHIC within perturbative QCD is an indirect
evidence for the saturation physics leading to the CGC. Note that,
implicit in these considerations, is the assumption that
the spectrum of the produced hadrons (as measured in the experiment)
can be identified with the  spectrum of the partons (mostly gluons)
liberated in the early stages of the collisions (as computed in
the CGC). This `duality' hypothesis is far from being
rigorous, but it has some theoretical motivation, and, more
importantly, it seems to be roughly consistent with the data
in various experimental settings. With this assumption, the
CGC prediction for the total
multiplicity ${\rm d}N/{{\rm d}\eta}$ at RHIC \cite{Raju}
turns out to be quite close to the actual
value measured in the experiments. In fact, simple calculations
within the CGC framework satisfactorily describe the
global features of the total multiplicity at RHIC,
like its dependencies upon the  energy $\sqrt{s}$ and the
rapidity $\eta$, and also upon the centrality of the collision
(or the ``number of participants'') \cite{KNL}.

Another interesting prediction of the CGC theory that will be
discussed at length in what follows refers to the interplay between the
energy dependence and the nuclear ($A$) dependence of the
{\it gluon spectrum} (by which I mean either the unintegrated
gluon distribution in the nuclear wavefunction, or the spectrum
of the gluons produced in a heavy ion collision, and which is 
further assimilated with the spectrum
${\rm d}N/{{\rm d}^2p_\perp\!{\rm d}\eta}$ of the produced hadrons).
Because of coherence effects associated with the non--linear
gluon dynamics, the gluon distribution
of a large nucleus is not simply the incoherent sum of the
corresponding distributions produced by $A$ separated nucleons.
This is true not only in the saturation region at low momenta
$p_\perp \simle Q_s$ (where, as we shall see, the gluon distribution
scales roughly like $A^{2/3}$ rather than like $A$), but also for momenta 
{\it well above} $Q_s$, where however the trend of the
$A$--dependence depends upon $y$ and
gets reversed when increasing the energy :
Whereas at low energies, the theory \cite{MV,K96,JKMW97,KM98} 
predicts an
{\it enhancement} of the $A$--dependence due to `higher--twist'
rescattering effects (which fall off rapidly with increasing
$p_\perp$; see Sect. III below), 
on the other hand, {\it at sufficiently large energies},
the quantum evolution of the color glass \cite{CGC} considerably
modifies the dominant, leading--twist, contribution
(this acquires an `anomalous dimension' \cite{GLR,SCALING,MT02,DT02}),
and thus provides a power--like {\it suppression} of the
$A$--dependence,
which persists up to relatively high $p_\perp$ \cite{SCALING}
(see also Sect. IV).
This suppression is further amplified by running coupling effects
\cite{AM03}.

This change of behaviour has dramatic consequences for the
spectrum of the produced gluons (or hadrons): At low energies,
one expects an {\it enhancement} at intermediate $p_\perp$ 
(``Cronin peak'') in the properly normalized production cross--section 
\cite{GJ02,JNV}, while at higher energies,
one should rather find a {\it depletion} 
within a wide range of momenta (``high--$p_\perp$ suppression'')
\cite{KLM02,KKT,Baier03,Nestor03}.

When the latter feature has been conceptually
first realized \cite{KLM02},
the ``high--$p_\perp$ suppression'' was freshly discovered at RHIC
\cite{QM02}  --- the high--$p_\perp$ yield
in gold--gold collisions was found to be almost an order
of magnitude lower than expected from jet production arising
from incoherent parton--parton scattering \cite{d'Enterria} 
---, so it was tempting to propose this ``quantum saturation'' 
scenario as a possible explanation of the
data \cite{KLM02}. To further substantiate this scenario,
it is worth mentioning that the onset of the
anomalous dimension is closely related to another important
prediction of the CGC, namely a new scaling law for the gluon
distribution \cite{SCALING,MT02,DT02,MP03},
which seems to be confirmed by the recent discovery of
``geometric scaling'' in the HERA data at small $x\le 10^{-2}$
\cite{geometric}. But in the context of Au+Au collisions at RHIC,
the argument in Ref. \cite{KLM02} suffers from a serious drawback:
A central--rapidity ($\eta=0$) jet with $p_\perp = 5 - 10$ GeV
is produced by partons with $x\sim 10^{-1}$, which
is a too large value of  $x$ for the quantum evolution of the CGC
to play
any role! Besides, the observed high--$p_\perp$ suppression can
be also understood as arising from a completely different physical
mechanism: energy loss through {\it final--state} interactions
(or `jet quenching') \cite{JetQuenching}.

\begin{figure}[t]
\centering
\includegraphics[height=0.30\textheight,width=.45\textwidth,clip]
{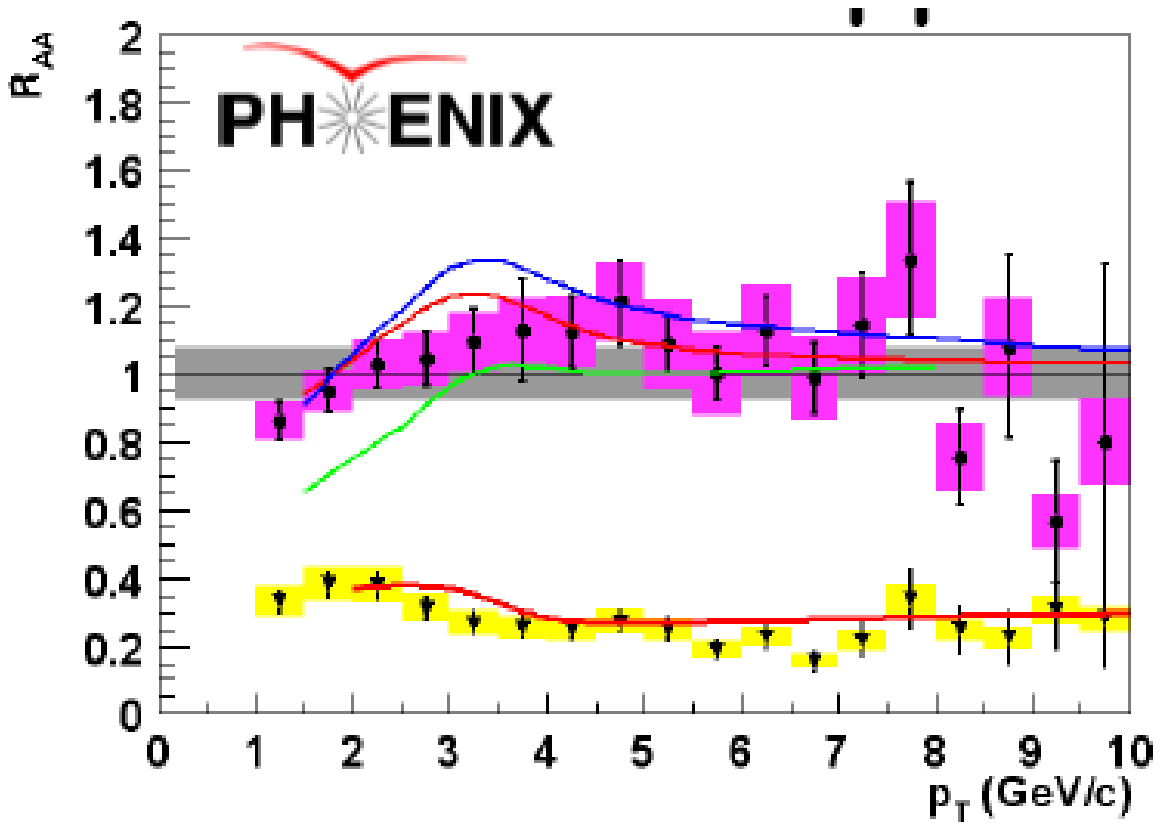}
\hspace*{.3cm}
\includegraphics[height=0.30\textheight,width=.45\textwidth,clip]
{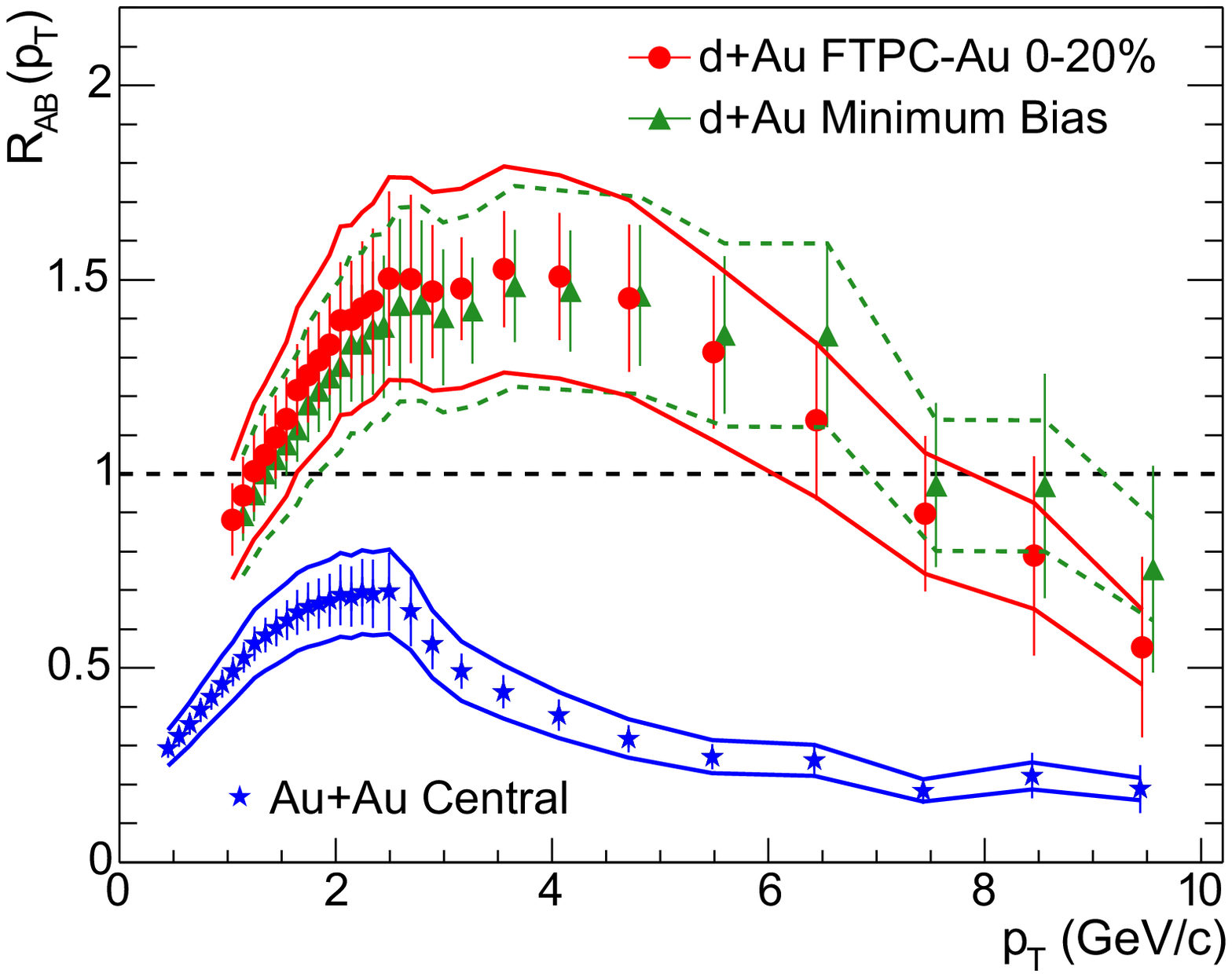}
\begin{minipage}[hbt]{6in}
\caption{{\small Central rapidity ($\eta=0$)
PHENIX $\pi^0$ and STAR $h^{\pm}$
data comparing $R_{\rm dAu}$ to $R_{\rm AuAu}$
~\protect{\cite{RHIC-dAu-mid}}.
These results, together with similar data from BRAHMS and 
PHOBOS~\protect{\cite{RHIC-dAu-mid}}, prove that jet quenching in Au+Au
collisions at central rapidity is necessarily a final state effect. }
\label{dAu0}}
\end{minipage}
\end{figure}

\begin{figure}[htbp]
\centerline{\includegraphics[height=4.cm]{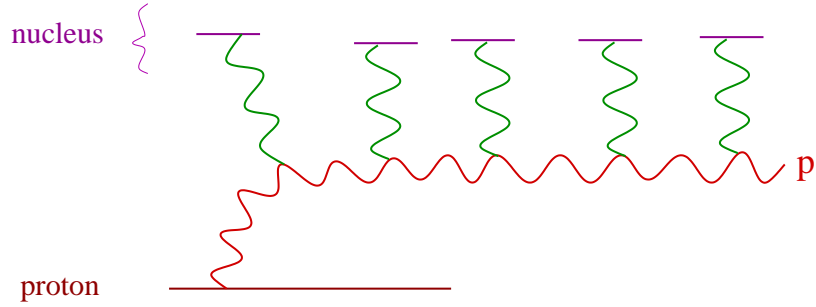}}
\begin{minipage}[hbt]{6in}
\caption{{\small Multiple scattering in a proton--nucleus collision.
In the experiments at RHIC, the proton is actually replaced by a deuteron.
}}
\label{dARESCAT1}\end{minipage}
\end{figure}

Nowadays, we know that this last explanation is in fact the
correct one. The decisive evidence in that sense came from another
set of experiments at RHIC in which one of the two heavy nuclei beams
has been substituted by a deuteron beam, thus eliminating the
final--state interactions. Fig. ~\ref{dAu0} exhibits the experimental
results obtained by PHENIX and STAR for the {\it nuclear modification
factor} in deuteron--gold collisions $(R_{\rm dAu})$ compared to
gold--gold $(R_{\rm AuAu})$. (BRAHMS and PHOBOS have reported
similar results; see Refs. in \cite{RHIC-dAu-mid}.)
Here, for generic nuclei A and B which scatter with
impact parameter $b_\perp$, the ratio $R_{\rm AB}$ is defined as:
\be\label{RAB}
   R_{\rm AB}( \eta, p_\perp,b_\perp)\,\equiv
\,\frac{1}{\langle N_{\rm coll}(b) \rangle}\,
   \frac{{\rm d}N_{\rm A+B}/ {\rm d}\eta{\rm d}^2p_\perp
{\rm d}^2b_\perp}
{{\rm d}N_{\rm p+p}/{\rm d}\eta{\rm d}^2p_\perp{\rm d}^2b_\perp
   }\,,
\ee
where
${\langle N_{\rm coll}(b_\perp) \rangle}$ denotes the average
number of collisions for a centrality bin at the given impact 
parameter\footnote{For `minimum bias' events which are averaged over
all values of $b_\perp$, ${\langle N_{\rm coll} \rangle}$  scales like
$AB/(R_A+R_B)^2$. E.g.: for A+A, ${\langle N_{\rm coll} \rangle}
\sim A^{4/3}$, while for d+A,  ${\langle N_{\rm coll} \rangle}
\sim A^{1/3}$.}.
This definition is such that $R_{\rm AB}$ would be equal to one if A+B
were an incoherent superposition of nucleon--nucleon collisions;
converserly, its deviation from one is
a measure of collective nuclear effects.
The data in Fig. ~\ref{dAu0} clearly show {\it enhancement} (rather
than suppression) in the $R_{\rm dAu}$ ratio at moderately large
$p_\perp$. As already mentioned, and will be detailed
in Sect. III, this {\it Cronin peak}
can be attributed to the multiple interactions suffered by the
colliding parton from the deuteron on its way through the
nucleus  (see Fig. ~\ref{dARESCAT1}).
The parton scatters off gluons with a relatively large momentum
fraction $x\sim 0.1$, but whose density is still quite large (hence
the importance of multiple scattering), because the gold
nucleus contains many color sources already at low energy:
the $3A\sim 600$ valence quarks. The results in
Fig. ~\ref{dAu0} demonstrate that the strong suppression
(by a factor 4 to 5) observed in $R_{\rm AuAu}$
cannot be attributed to the quantum evolution
of the gluon distribution in the nuclear wavefunction
(otherwise, a similar effect would be seen at $\eta=0$
also in  $R_{\rm dAu}$), and thus
indirectly confirm its origin as a final--state effect.

\begin{figure}[htbp]
\includegraphics[height=5.4cm]{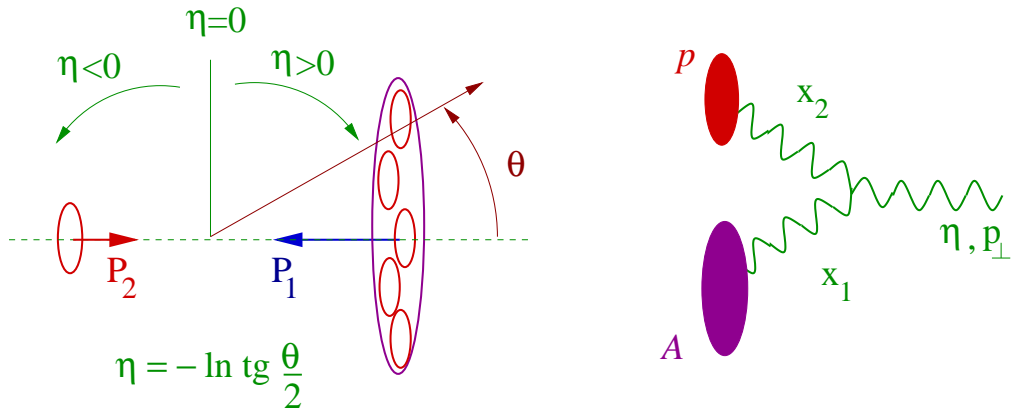}
\begin{minipage}[hbt]{6in}
\caption{{\small Kinematics for gluon production
via gluon fusion in proton($p$)--nucleus($A$) collisions
(d+Au collisions are similar). With light--cone notations
[$k^\mu=(k^+,k^-,{\bf k}_\perp)$, with $k^\pm = (E_k\pm k_z)/ \sqrt{2}$],
the nucleons involved in the collision have purely longitudinal
momenta: $P^\mu_1=(0,P_1,0_\perp)$ and $P^\mu_2=(P_2,0,0_\perp)$,
with $P_1=P_2=\sqrt{s/2}$ in the center-of-mass frame, while 
the emerging gluon has $p^\mu=\Big(
\frac{p_\perp}{\sqrt{2}}\,{\rm e}^\eta,
\frac{p_\perp}{\sqrt{2}}\,{\rm e}^{-\eta},
{\bf p}_\perp\Big)$. 
Energy--momentum conservation implies $
x_1 \equiv {k_{1}^-}/{P_1} =({p_\perp}/{\sqrt{s}})\,{\rm
e}^{-\eta}$ and $x_2\equiv
{k_{2}^+}/{P_2}=(p_\perp/{\sqrt{s}})\,{\rm e}^{\eta}$,
where $k_{1,2}$ refer to the two gluons which fuse with each other.
Thus, larger positive values for $\eta$ correspond to smaller
values for the longitudinal fraction  $x_1$ of the gluon from the nucleus.
}}
 \label{KINEM}\end{minipage}
\end{figure}

\begin{figure}[htb]
  \centering
\includegraphics[height=4.5cm]{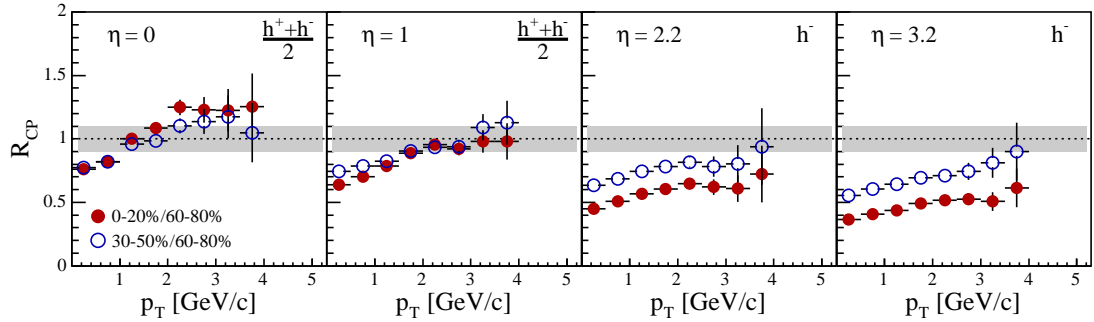}
\begin{minipage}[hbt]{6in}
        \caption{{\small  d+Au collisions at
BRAHMS~\protect{\cite{Brahms-data}}
: The ratio $R_{\rm CP}$ of yields from
collisions of a given centrality class (0-20\% or 30-50\%) to yields
from more peripheral collisions (60-80\%), scaled by the mean number
of binary collisions in each sample, as a function of $p_\perp$
for central and forward pseudorapidities.}
        \label{brahmsCP}}\end{minipage}
\end{figure}

More recently, the BRAHMS experiment has presented
the first results for $R_{\rm dAu}$ at `forward rapidities'
($\eta=2-3$ in the deuteron fragmentation region; see
Fig.~\ref{KINEM}) \cite{Brahms-data}, which have been soon after
confirmed by the other collaborations \cite{QM04}. The most
remarkable feature about these new data is that they
show a strong dependence upon $\eta$, and also a general trend 
with both $\eta$ and the centrality
which at a first sight looks counterintuitive and surprising:
As argued before, the Cronin enhancement is attributed
to multiple scatterings between a parton from the deuteron
and the gluons in the nucleus. When increasing $\eta$,
we are probing gluons with smaller values of $x$
(see kinematics in the capture to Fig.~\ref{KINEM}), so the gluon
distribution must increase,
and this should favour the rescattering effects and thus enhance
the Cronin peak. Similarly, more central collisions are probing
regions with higher density, which should again enhance the Cronin
peak. But the experimental results show precisely
the opposite trends (cf. Fig. ~\ref{brahmsCP}) !

 \begin{figure}[htbp]
 \centering
\includegraphics[height=4.5cm]{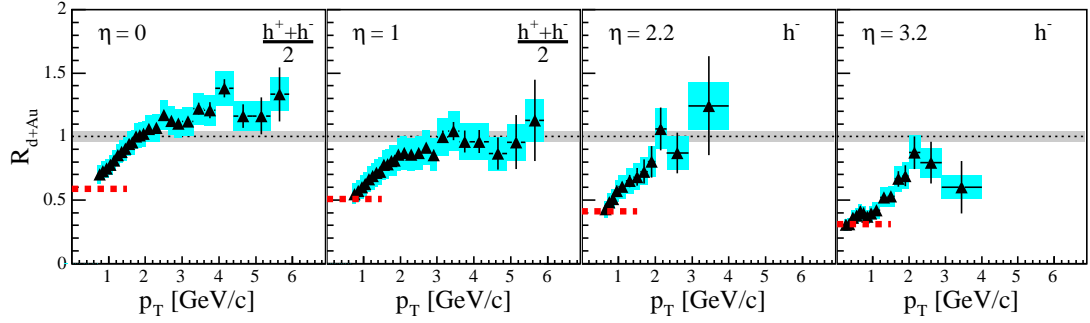}
\begin{minipage}[hbt]{6in}
        \caption{{\small BRAHMS results for d+Au collisions at
 200 GeV/nucleon \protect{\cite{Brahms-data}} : The ratio
$R_{\rm dAu}$ for charged hadrons as a function of $p_\perp$
for central and forward pseudorapidities.}}
 \label{brahmsR}\end{minipage}\end{figure}

But as counterintuitive as it might seem, this
behaviour has been in fact anticipated, on the basis of the CGC
ideas \cite{KLM02,KKT,Nestor03} : As already mentioned, one effect
of the quantum evolution towards smaller values of $x$ is to suppress
the dependence of the high--$p_\perp$ spectrum upon the atomic number
$A$, or, more generally, upon the number of colliding particles in a given
centrality bin. What came nevertheless as a surprise
is the fact that this suppression sets in {\it so fast} !
As manifest on Fig. ~\ref{brahmsCP}, the  centrality
dependence gets reversed already for $\eta=2$
(which corresponds to $x\sim 10^{-3}$ for 
$p_\perp = 4$ GeV), while for $\eta=3.2$ ($x\sim 5\times 10^{-4}$),
the data in Fig. ~\ref{brahmsR} show suppression
($R_{\rm dAu} < 1$) at all the measured values of $p_\perp$.

In fact, already  before the advent of the BRAHMS data,
the rapid evolution of the ratio
$R_{\rm AB}$ with increasing $\eta$ has been observed 
theoretically \cite{Nestor03},
in a numerical calculation based on the Balitsky--Kovchegov (BK)
equation \cite{B,K}. In Ref. \cite{Nestor03}, the Cronin peak 
was seen to flatten out and completely disappear after
only one unit of rapidity. But the mechanism
responsible for such a rapid evolution has been fully elucidated
only subsequently, through an exhaustive analytic study \cite{IIT04},
which has also clarified other important issues like
the respective roles of the proton and the nucleus in the evolution 
with $\eta$, and the effects of the running of the coupling. The main
results of this study will be briefly described in the remaining
part of this review.


\section{II. The ${R}_{\rm pA}$--ratio in the CGC}

For simplicity, in what follows I shall discuss proton($p$)+nucleus($A$)
scattering, and assume that similar results hold
also for the d+Au collisions of interest at RHIC. The physically
interesting quantity is then the ratio (compare to Eq.~(\ref{RAB}))
\be\label{RpA}
   R_{\rm pA}(\eta,p_\perp)\,\equiv
\,\frac{1}{A^{1/3}}\,
   \frac{\frac{{\rm d}N_{\rm pA}}{{\rm d}\eta{\rm d}^2p_\perp
{\rm d}^2b_\perp}}
 {\frac{{\rm d}N_{\rm pp}}{{\rm d}\eta{\rm d}^2p_\perp{\rm d}^2b_\perp
   }}\,,
\ee
between the yield of produced gluons per unit {\it phase--space}
(i.e., per unit of pseudorapidity, transverse momentum and impact parameter)
in $pA$ and $pp$ collisions normalized by $A^{1/3}$ (since, at a fixed
impact parameter, the number of nucleons available for scattering
in a $pA$ collision is larger by a factor $A^{1/3}$ than in the
corresponding $pp$ collision; note that I assume homogeneity
in impact parameter space, for simplicity). Thus, the ratio (\ref{RpA})
measures the difference between $pp$ collisions and an incoherent
superposition of proton--nucleon collisions, and is an useful
observable to pinpoint collective effects like gluon saturation
in the wavefunction of the incoming nucleus
\footnote{Recall that, in $pA$ collisions,
we do not expect this ratio to be modified by final--state interactions.}.

So, how to compute gluon production in $pA$ collisions ?
Leaving aside the fully numerical calculations \cite{Raju,JNV} 
(and Refs. therein)
based on the classical version of the CGC theory (the 
McLerran--Venugopalan model \cite{MV,CGCreviews}; see also Sect. III below),
which apply to both $pA$ and $AA$ collisions but have not been
extended yet to include the quantum evolution with $\eta$,
all (semi)analytic calculations of the ${R}_{\rm pA}$--ratio
\cite{KLM02,KKT,Baier03,Nestor03,IIT04} rely on the factorization of
the cross--section for particle production in terms of
the gluon distributions in the target and the projectile
(``$k_T$--factorization''), which has been since long proven for
$pp$ collisions (see, e.g., \cite{FR}), 
and has been shown in the recent years to also
hold for $pA$ collisions \cite{KM98,DM01,KT02,BGV04},
provided a special definition is used
for the nuclear gluon distribution \cite{Braun}
which incorporates rescattering effects.

Up to an irrelevant
normalization factor which cancels out in the ratio (\ref{RpA}),
the gluon yield in $pA$ collisions can be computed as (see
kinematics in Fig.~\ref{KINEM})
\be\label{dNpA}
\frac{{\rm d}N_{\rm pA}}{{\rm d}\eta{\rm d}^2p_\perp
{\rm d}^2b_\perp}\,\propto\,
\frac{1}{p_\perp^2} \int
d^2k_\perp\,\varphi_A(k_\perp,y_1)\,\varphi_p(p_\perp-k_\perp,y_2)\,,\ee
where $y$ denotes the gluon rapidity, related to its
longitudinal momentum fraction $x$ via $y=\ln
(1/x)$, and $\varphi_A(\kk,y)$ is the {\it gluon occupation factor}
\footnote{I am glossing here over the subtle (but not essential
for the present discussion) distinction between the {\it canonical}
gluon occupation factor (defined as the expectation value
of the Fock space number operator $a^\dagger_k a_k$ in some specific
gauge \cite{CGCreviews,SCALING}), which is the quantity meant
by Eq.~(\ref{phidef}), and the special `gluon distribution'
(a scattering amplitude for a color dipole) which strictly speaking enters
the $k_T$--factorized expression (\ref{dNpA}) for the gluon production
(see, e.g.,  \cite{Braun,BGV04} for details).},
i.e., the number of gluons of given spin and color per unit
phase--space in a nucleus with atomic number $A$:
\be\label{phidef}
\varphi_A(\kk,y)\,\equiv\,\frac{(2\pi)^3}{2(N_c^2-1)}\,
\,\frac{{\rm d} N^{\rm gluon}_A}{{\rm d} y \,{\rm d}^2k_\perp{\rm
d}^2 b_\perp}\,.\ee ($N_c=3$ is the number of `colors'.)
For $pp$ collisions, a similar formula holds with $\varphi_A(k_\perp,y_1)$
replaced by $\varphi_p(k_\perp,y_1)$. Note
that $y_{1,2} = y_0 \pm \eta$, with $y_0\equiv \ln({\sqrt{s}}/k_\perp)$
(cf. Fig.~\ref{KINEM}). Thus increasing $\eta$ amounts to
increasing $y_1$ but decreasing $y_2$.

Clearly, the nuclear effects enter the previous formulae via the nuclear
gluon distribution $\varphi_A$ : Unlike its proton counterpart
$\varphi_p$, which is the standard, leading--twist, `unintegrated
gluon distribution' generally used in conjunction with $k_T$--factorization
\cite{FR}, $\varphi_A$ encodes higher--twist effect associated with
the rescattering suffered by the gluon partaking in the collision
while it crosses the nucleus.

At central pseudo--rapidity ($\eta=0$)
and for the high--$p_\perp$ kinematics at RHIC, $y_1=y_0$ is rather
small, so one can assume that the gluons in the nuclear distribution
$\varphi_A(\kk,y_1)$ are produced by direct radiation from
the valence quarks. In that case, Eq.~(\ref{dNpA})
describes the multiple scattering between a gluon in the proton
and the valence quarks in the nucleus,
as illustrated in Fig.~\ref{dARESCAT1}. The effect of the
rescattering on the ratio ${R}_{\rm pA}$ 
will be discussed in the next section, within the context of a simple model 
for the distribution of the valence quarks, introduced by McLerran and
Venugopalan \cite{MV}.

\begin{figure}
\centerline{\includegraphics[height=5.cm]{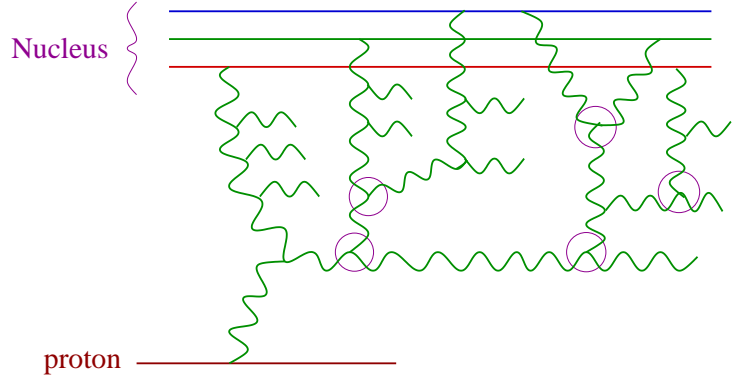}}
\caption{{\small $pA$ collision at forward rapidity: the
(relatively large--$x$) gluon emitted by the incoming proton scatters off
the highly--evolved gluon distribution in the nuclear
wavefunction at small--$x$. Non--linear effects (which correspond
to either gluon recombination in the nuclear wavefunction, or multiple
scattering of the `external' gluon) are indicated by small cercles.
}}\label{PHI_EVOLVE}
\end{figure}
At forward, and sufficiently large,
pseudo--rapidities ($\eta=2-3$), the gluons with $y_1= y_0 +\eta$
in the nuclear wavefunction are predominantly produced by quantum
evolution from lower values of $y_1$ \cite{BFKL},
that is, they are emitted by gluons with larger values of $x$
which are themselves radiated by the valence quarks.
Since the gluon density was rather large to start with (due to
the large number of valence quarks), and it is further amplified
by the evolution with $\eta$, the newly emitted gluons propagate
in a strong background field --- the color field created by sources 
(valence quarks or gluons) at larger values of $x$ --- and
undergo multiple scattering off the latter (see Fig. \ref{PHI_EVOLVE}).
This rescattering introduces a negative feedback (it reduces the
number of gluons), and eventually leads to gluon saturation.
Computing $\varphi_A$ in this non--linear
regime requires an analysis of the quantum evolution in the presence
of strong fields, which has been completed only recently
\cite{CGC,B,K,JKLW97,W,SAT,AM01,SCALING,MT02}, 
with results to be discussed in Sect. IV.

The simplify the discussion while keeping all the salient
physical features, it is convenient to replace the ratio
in Eqs.~(\ref{RpA})--(\ref{dNpA}) by the simpler quantity
\be\label{Rdef} {\mathcal
R}_{pA}(k_\perp,y)\,\equiv\,\frac{\varphi_A(k_\perp,y)}{A^{1/3} \,
\varphi_p(k_\perp,y)}\,,\ee
to be considered for $y=y_2$, i.e., for the rapidity of the gluon
in the nuclear wavefunction which participates in the collision.
Eq.~(\ref{Rdef}) measures the difference between the gluon
distribution in the nucleus and that in the proton (scaled up by
$A^{1/3}$) at the same (small) value of $x$, and thus is the most
direct expression of the nuclear effects which should be responsible
for the deviation of the experimentally measured ratio ${R}_{\rm pA}$
from one. In particular, qualitative effects like the Cronin enhancement
or the high--$p_\perp$ suppression should be already visible (and
theoretically easier to study) on the ratio (\ref{Rdef}); this
is confirmed by the analyses in Refs. \cite{KKT,Nestor03},
which have found a similar qualitative behaviour for the quantities
defined in Eqs.~(\ref{RpA})--(\ref{dNpA}) and, respectively,
Eq.~(\ref{Rdef}). In what follows I shall exclusively focus
on the simpler ratio, Eq.~(\ref{Rdef}).

\section{III. Central rapidity: Cronin peak in the MV model}

So long as the rapidity $y$ is not too large,
one can ignore quantum evolution towards small $x$
and describe the gluon distribution in the nucleus as the result
of classical radiation from the valence quarks. This is probably a good
approximation for particle production in d+Au collisions at central
rapidity ($\eta=0$), since in that case $y=y_0 \sim 3$ is still quite
small. (From experience with the phenomenology at HERA, we expect
small--$x$ evolution effects to become important only for $y>5$,
corresponding to $x < 10^{-2}$ ; see e.g. \cite{IIM03} and
references therein.)

The gluon spectrum
${\rm d} N_A/{\rm d} y \,{\rm d}^2k_\perp{\rm d}^2 b_\perp$
is produced by the valence quarks located within a tube of transverse
area $\sim 1/k_\perp^2$ which traverses the nucleus in longitudinal
direction at impact parameter $b_\perp$ (so the length of the tube
scales like $A^{1/3}$). If the transverse section of this tube
is much smaller than the area of a single nucleon (i.e.,
$k_\perp^2\gg \Lambda_{\rm QCD}^2$), then the valence quarks which
are included inside the tube belong typically to different nucleons.
It is then reasonable to assume, following McLerran and Venugopalan
\cite{MV}, that these quarks are {\it uncorrelated} with each other.
But although the {\it color charges} of these quarks sum up
incoherently, this is generally not so for the corresponding gluon
distributions (unlike it would happen for photon distributions in
QED), because the classical color fields obey a
{\it non--linear} equation: the Yang--Mills equation.

It is convenient to consider first the case of a proton,
since there the color fields are still weak, and one can rely
on the linear  approximation. Then, the total
gluon distribution is simply the incoherent sum of the distributions
produced by the $N_c=3$ valence quarks, which by themselves can
be evaluated with the well known formula for bremsstrahlung:
\be\label{bremss}
\frac{{\rm d} N_{\rm quark}}{{\rm d} y \,{\rm d}^2k_\perp}
\,=\,\frac{\alpha_s C_F}{\pi^2}\,\frac{1}{k_\perp^2}\,
.\qquad\quad
(C_F=(N_c^2-1)/2N_c)\ee
After multiplying this by $N_c$ and dividing the result by the number of
(gluonic) color states $N_c^2-1$ and by the proton area $\pi R^2_p$,
one finds the occupation factor as (cf. Eq.~(\ref{phidef})):
\be\label{phip0}
\varphi_A(\kk)\,\simeq\,\frac{\mu_p}{k_\perp^2}\,,\qquad
\mu_p\,\equiv\,\frac{2\alpha_s}{R^2_p}\,.\ee
Physically, $\mu_p$ is the color charge squared of
the valence quarks per unit transverse area (here, in the proton).
Eq.~(\ref{phip0}) is valid for $\kk\gg Q_p$ with
$Q_p\sim \Lambda_{\rm QCD}$ a non--perturbative scale related
to confinement.

Moving to a large nucleus, one finds $\mu_A\simeq A^{1/3}\mu_p$, but
the calculation of the corresponding gluon distribution is more elaborate,
as it requires the exact solution to the Yang--Mills equation
with a strong color source. This has been solved indeed,
with the following result \cite{JKMW97,K96,KM98,CGCreviews} :
 \be\label{phiMV} \varphi_A(\kk)\,=\, \!\int \!
d^2r_\perp \,{\rm e}^{-ik_\perp\cdot r_\perp}\,\, \frac{1-{\rm
exp}\Big\{-\frac{1}{4}\,\alpha_s N_c  r_\perp^2 \mu_A \ln{4\over
r_\perp^2\Lambda^2}\Big\}}{\pi \alpha_s N_cr_\perp^2 }\,,\,\,\ee
where $\Lambda\sim \Lambda_{\rm QCD}$ is a non--perturbative scale
introduced by hand to cut off an infrared
divergence which physically should be removed by confinement.
This formula features the {\it saturation momentum} $Q_s(A)$,
defined by
\be\label{QsatMV}
Q_s^2(A)\,=\,\alpha_s N_c \mu_A\,\ln\frac{Q_s^2(A)}{\Lambda^2} \,\,\sim\,
A^{1/3}\,\ln A^{1/3}\,,\ee
(note that when $r=2/Q_s(A)$ the exponent in Eq.~(\ref{phiMV})
is equal to one),
as the intrinsic transverse momentum scale which separates
the linear regime from the non--linear one:

(i) At high momenta $k_{\perp} \gg Q_s(A)$, one can expand the
exponential in Eq.~(\ref{phiMV}), and thus obtain (I only show
here the first two terms in this expansion) :
\be\label{phiA0}
\varphi_A(\kk)\,\simeq\,\frac{\mu_A}{k_\perp^2}\,+\,\alpha_s N_c
\left(\frac{\mu_A}{k_\perp^2}\right)^2\left[\ln\frac{k_\perp^2}{\Lambda^2}
+ 2\gamma-2\right],\qquad {\rm for\quad}k_{\perp} \gg Q_s(A)
\,.\ee
The first term, which
gives the dominant behaviour at  high $k_\perp$, is recognized as
the bremsstrahlung spectrum, corresponding to independent
emissions from the valence quarks. This is as
expected: At high momenta, non--linear
effects become negligible because the corresponding modes have
only little overlap with each other. Thus, for sufficiently high
momenta, $\varphi_A(\kk)\,\simeq\, A^{1/3}\varphi_p(\kk)$, and
the ratio ${\mathcal R}_{pA}(k_\perp)$ approaches one {\it from 
the above}.

(ii) At low momenta $k_{\perp} \ll Q_s(A)$ (with $\kk\gg \Lambda$
though), the dominant behavior is obtained after
neglecting the exponential term in Eq.~(\ref{phiMV}), and reads
\be\label{phiAlow}
\varphi_A(\kk) \,\simeq\,
\frac{1}{\alpha_s N_c}\,\left\{\ln \frac{Q_s^2(A)}{k_{\perp}^2}
\,+\,\mathcal{O}\big(1\big)\right\}\,,
\qquad {\rm for\quad}k_{\perp} \ll Q_s(A)\,.\ee
This is parametrically large, $\mathcal{O}\big(1/\alpha_s)$,
but increases only slowly (logarithmically) when decreasing $k_\perp$
and/or increasing the atomic number $A$. This behaviour, which should
be contrasted with the corresponding power--law behaviour (in both
$1/k_\perp^2$ and $A$) of Eq.~(\ref{phiA0}), is the hallmark of
{\it gluon saturation} (here, within the MV model).

Because of this change in behaviour, 
the ratio (\ref{Rdef}) becomes very small at low momenta 
$k_{\perp} \ll Q_s(A)$ (where it is proportional to $k_\perp^2$), 
but it increases rapidly with $z\equiv \kk^2/Q_s^2(A)$, and becomes {\it
parametrically large}
(and, in particular, larger than one) for momenta in the vicinity
of the nuclear saturation momentum. Indeed, for $z\sim 1$,
$\varphi_A\sim 1/\alpha_s N_c$ and $\varphi_p\sim\mu_p/Q_s^2(A)$
(cf. Eq.~(\ref{phip0})), and by also using $\mu_A = A^{1/3}\mu_p$,
one finds
 \be\label{RmaxMV}
 {\mathcal R}_{pA}(k_\perp)\,\sim\,\rho_A\,>\,1\,
\qquad {\rm for\quad}k_{\perp} \sim Q_s(A)\,,\ee where (cf.
Eq.~(\ref{QsatMV})) : \be\label{rhoA} \rho_A\,\equiv
\,\frac{Q_s^2(A)}{\alpha_s N_c \mu_A}\,=\,
 \ln\frac{Q_s^2(A)}{\Lambda^2}\,\sim \ln A^{1/3}\,.\ee
The estimate (\ref{RmaxMV}) holds up to corrections of
$\mathcal{O}(1)$, which are indeed suppressed in the limit
of a very large nucleus, for which $\rho_A\gg 1$. To make contact
with RHIC phenomenology, it is useful to notice that, for a
gold nucleus at RHIC energies, one expects $Q_s^2(A)\simeq 2$
GeV$^2$ \cite{GM04}, which together with $\Lambda \simeq 200$ MeV implies
$\rho_A\simeq \ln 50 \simeq 4$.

\begin{figure}[t]
\centering
\includegraphics[height=0.3\textheight,width=.48\textwidth,clip]
{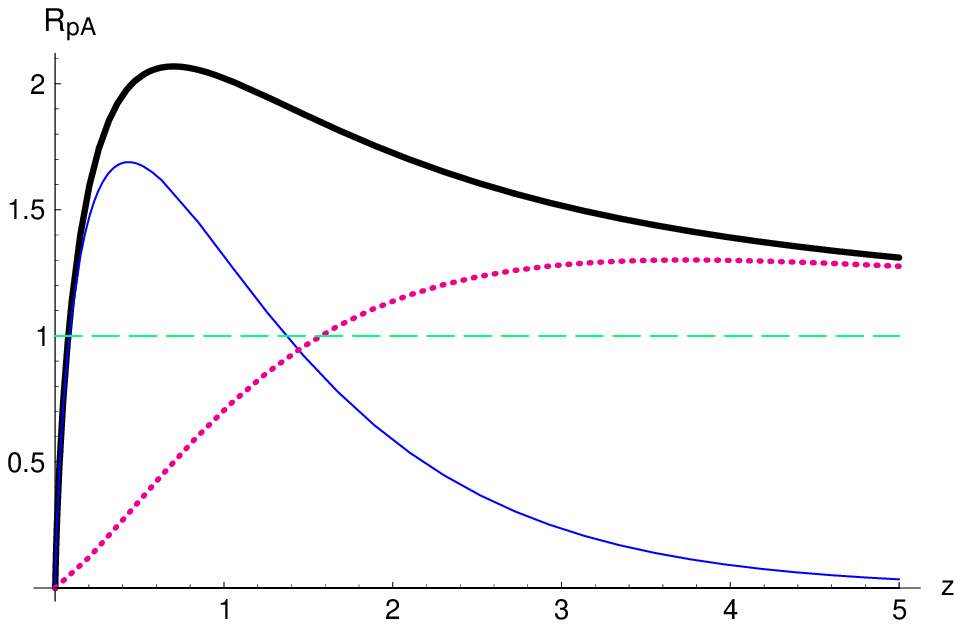}
\hspace*{.3cm}
\includegraphics[height=0.3\textheight,width=.48\textwidth,clip]
{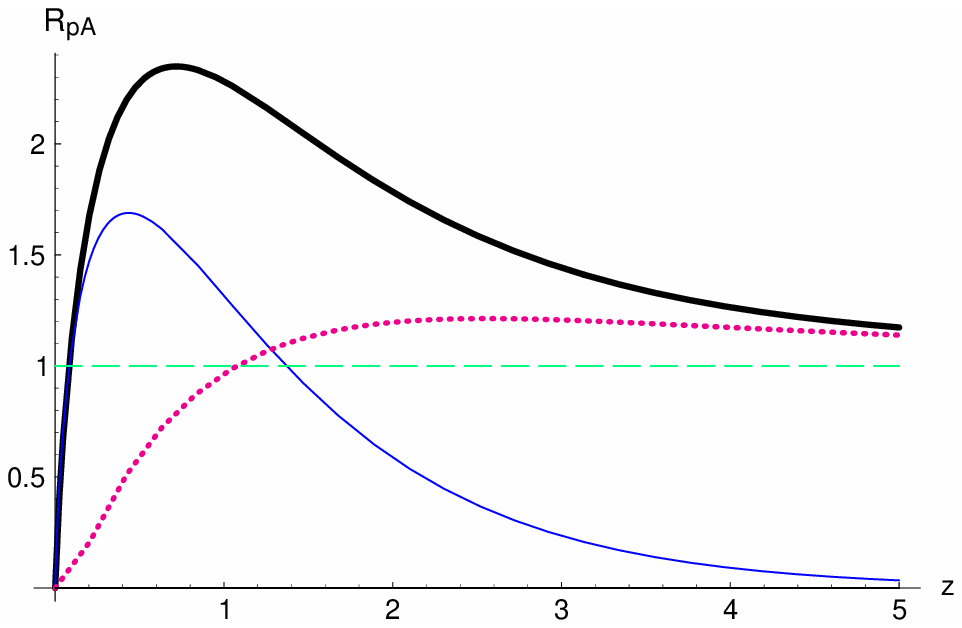}
\begin{minipage}[hbt]{6in}
\caption{{\small The ratio
    $\mathcal{R}_{pA}(z)$ as a function of the scaled momentum
    variable $z=k^2/Q_s^2(A)$ in the fixed (left) and running
    (right) coupling McLerran-Venugopalan model for $\rho_A=6$.}
    {\small The thick line corresponds the ratio
    $\mathcal{R}_{pA}(z)$; the solid line shows the
    saturation contribution $\mathcal{R}^{\rm sat}_{pA}(z)$; the
    dotted line shows the twist contribution
    $\mathcal{R}^{\rm twist}_{pA}(z)$. (From Ref. \cite{IIT04}.)}}
\label{RpAMV}
\end{minipage}
\end{figure}

The previous considerations show that the ration ${\mathcal
R}_{pA}$ {\it must have a maximum} as a function of $k_\perp$,
with the position of the maximum near $Q_s(A)$ and the height of
the peak of $\mathcal{O}\big(\ln A^{1/3}\big)$. This is confirmed
by an exact (analytic) calculation \cite{IIT04} of the nuclear
gluon spectrum in the MV model, Eq.~(\ref{phiMV}), which together
with Eq.~(\ref{phip0}) for the corresponding spectrum in the
proton leads to the ratio ${\mathcal R}_{pA}$ displayed in
Fig.~\ref{RpAMV} (the left figure there). The location $z_0$ of
the maximum and its height ${\mathcal R}_{\rm max}(A)\equiv
{\mathcal R}_{pA}(z_0)$ can be computed analytically, in an
expansion in powers of $1/\rho_A$ \cite{IIT04} :
\be\label{RmaxMVf} z_0\,=\,
0.435\,+\,\frac{0.882}{\rho_A}\,+\,{\mathcal
O}\big(\rho_A^{-2}\big),\qquad {\mathcal R}_{\rm
max}(A)\,=\,0.281\,\rho_A\,+\,0.300\,+\,{\mathcal
O}\big(\rho_A^{-1}\big)\,.\ee

In Fig.~\ref{RpAMV}, it is also shown the result of a calculation
including {\it running coupling} effects in the MV model\footnote{The
inclusion of running coupling effects in an otherwise classical calculation
may look as rather ad--hoc. Still, this has the merit to
approximately include an important class of quantum corrections,
which are potentially large.} (the right figure there). This is obtained
by simply replacing the explicit factors of $\alpha_s$ in Eq.~(\ref{phiMV})
by the one--loop running coupling
: $\alpha_s\rightarrow
\alpha_s(4/r_\perp^2)$, with $\alpha_s(Q^2)\equiv
b_0/\ln(Q^2/\Lambda_{\rm QCD}^2)$. Then, the scale which
plays the role of a saturation momentum reads simply
$Q^2_s(A)\equiv b_0N_c \mu_A$, and scales like $A^{1/3}$.
In that case too, the gluon distribution in the MV model can be
computed analytically  \cite{IIT04}, and the results are rather
similar (at least, qualitatively) to the fixed coupling case.
In particular, the parameter $\rho_A \equiv \ln{Q_s^2(A)}/{\Lambda^2}$
plays again an important role --- with a running coupling, this
appears even more naturally, as the inverse of
the running coupling: $\rho_A = b_0/\alpha_s(Q^2_s(A))$ ---, and
expansions like those in Eq.~(\ref{RmaxMVf}) can be again derived.
(It turns out that the coefficients of the dominant terms
in these expansions --- for both $z_0$ and  ${\mathcal
R}_{\rm max}(A)$ --- are the same for fixed and running
coupling.)

The plots in Fig.~\ref{RpAMV} not only confirm that a maximum
exists, but also show that this maximum is rather
{\it well pronounced}, which could not have been anticipated
solely on the basis of the previous approximations,
Eqs.~(\ref{phiA0}) and (\ref{phiAlow}). To understand that, one
needs also the behaviour of the nuclear gluon spectrum $\varphi_A(\kk)$
for momenta around $Q_s(A)$. The calculations in Refs. \cite{IIT04,Adrian}
reveal that, when increasing $\kk$ above  $Q_s(A)$, $\varphi_A(\kk)$
starts by decreasing {\it exponentially} with $z=k^2/Q_s^2(A)$,
before eventually relaxing, for $z > \ln\rho_A$,
to the power law decay displayed in Eq.~(\ref{phiA0}).
To better illustrate this, Fig.~\ref{RpAMV} exhibits also the
individual contributions to $\mathcal{R}_{pA}$ denoted as
$\mathcal{R}^{\rm sat}_{pA}$ and $\mathcal{R}^{\rm twist}_{pA}$
which are obtained after separating $\varphi_A(\kk)$ into two pieces:
the `twist' piece $\varphi_A^{\rm twist}$, which resums all the
terms which at high--momenta decay as inverse powers of $z$
(the first two terms in this series are shown in Eq.~(\ref{phiA0})),
and the `saturating' piece $\varphi_A^{\rm sat}\equiv \varphi_A
- \varphi_A^{\rm twist}$, which at $z\ll 1$ has the
logarithmic behaviour shown in Eq.~(\ref{phiAlow}), whereas
for $z>1$ decays exponentially with $z$.
For any $z\simle \ln\rho_A$, the saturating piece dominates over
the twist piece, and thus $\mathcal{R}_{pA}$ has a rapid fall off
at momenta just above the maximum (within the range
$1\simle z \simle \ln\rho_A$), leading to the well--pronounced
Cronin peak manifest in Fig.~\ref{RpAMV}.

To conclude this discussion, let me comment on the physical origin
of the Cronin enhancement in this classical model for saturation.
As already mentioned in the Introduction, the Cronin peak is generally
associated with multiple scattering within the nucleus
(cf. Fig. ~\ref{dARESCAT1}). This is also
the case in the present context, except for the fact that, since
we are looking directly at the wavefunction (rather than at a
scattering process), the multiple scattering is replaced by non--linear
effects in the production of virtual gluons from the valence quarks.
(One can say that, after one gluon is produced by some quark, it
scatters off the color fields produced by the other quarks.)
Since the gluon mutual interactions are repulsive, and are stronger
at low momenta, their net effect is to rearrange the gluon distribution
in transverse momentum space \cite{KKT,IIT04}:
Whereas in the bremsstrahlung spectrum
$\propto 1/\kk^2$ most quanta are located at low momenta, in
the presence of non--linear effects, the would--be low--$\kk$
gluons are pushed towards higher momenta, in such a way to minimize
their repulsion. Some of these `displaced' gluons are responsible
for the `higher--twist' contributions to the tail of the distribution
at $\kk \gg Q_s(A)$ (cf. Eq.~(\ref{phiA0})). But most of them
are quasi--uniformly redistributed at lower momenta $\simle Q_s(A)$,
thus giving a saturation plateau at $\kk\le Q_s(A)$
(cf. Eq.~(\ref{phiAlow})) \cite{IIT04}.

\section{IV. Non--linear gluon evolution in the CGC}

The non--linear evolution of the gluon distribution with increasing $y
=\ln 1/x$ is described by the renormalization group equation for the CGC,
also known as the JIMWLK equation \cite{CGC,JKLW97,W}, which is a
functional (or operator) equation, that is, it generates an infinite
hierarchy of ordinary integro--differential equations which couple
the evolution of various $n$--point functions. These equations resum
those radiative corrections which are enhanced by either the large
rapidity gap (namely, the terms of order $(\alpha_s y)^n$ for
any $n\ge 1$), or the high gluon density (since, e.g., the gluon
occupation factor being of $\mathcal{O}\big(1/\alpha_s)$ at saturation,
it interferes with the perturbative expansion). This complicated evolution
is illustrated in Fig. \ref{PHI_EVOLVE} which shows a typical 
gluon cascade which develops in the nuclear wavefunction at small $x$.
The horizontal rungs in this cascade are radiated gluons which are
strongly ordered in rapidity, while the mergings between various vertical
branches, as well as the multiple scattering of the produced gluon,
are representative for the non--linear effects in this high--energy
environment.  The general solution to the JIMWLK equation is not known,
but approximate solutions have been constructed which separately cover
the {\it non--linear regime} deeply at saturation, $\kk\ll Q_s(A,y)$,
(where remarkable simplifications occur in the evolution equations, due to
saturation \cite{SAT}), and the {\it linear regime} at $\kk\gg Q_s(A,y)$,
where the gluon occupation factor $\varphi_A(\kk,y)$ 
obeys a closed, linear, equation: the BFKL equation \cite{BFKL}.

Remarkably, it turns out that the gluon occupation factor deeply
at saturation ($\kk\ll Q_s(A,y)$) has the same simple form as in
the classical MV model (cf. Eq.~(\ref{phiAlow})), namely,
\be\label{phiAsat} \varphi_A(\kk,y)\,\approx\,
 \frac{1}{\alpha_s N_c}\,
\left\{\ln \frac{Q_s^2(A,y)}{k_{\perp}^2}
\,+\,\mathcal{O}\big(1\big)\right\}\,,
\qquad {\rm for\quad}k_{\perp} \ll Q_s(y,A)\,,\ee
where the saturation momentum is exponentially increasing with $y$
\cite{GLR} : 
\be\label{Qsfix0}
Q_s^2(A,y)\,\simeq\,Q_s^2(A)\, {\rm e}^{c\bar\alpha_s y}
\,, \qquad  {\rm (fixed\,\,coupling)}, \ee
where $c\simeq \,4.88$ and $\bar\alpha_s\equiv \alpha_s N_c/\pi$.
In this and the subsequent formulae, the
variable $y$ denotes the {\it difference} from
the rapidity $y_0$ corresponding to $\eta=0$; 
that is, from now on,  $y$ is numerically
the same as the pseudo--rapidity $\eta$.

Note the special way how Eq.~(\ref{phiAsat}) depends upon
$A$, $y$ and $k_{\perp}$ : $\varphi_A(\kk,y)$
is solely a function of the dimensionless ratio
$z\equiv k_{\perp}^2/Q_s^2(A,y)$. This
property, known as {\it geometric scaling} \cite{geometric},
reflects the fact that $Q_s(A,y)$ is the only
intrinsic scale at saturation.

Another remarkable feature is that geometric scaling is
approximately preserved within a wide range of momenta above
$Q_s$, where the evolution is linear \cite{SCALING}. One finds
indeed that, within the range \be\label{swindow1}
Q_s(A,y)\,\ll\,k_{\perp}\,\ll\,Q_s^2(A,y)/Q_s(A) \qquad  {\rm
(`extended\,\,scaling\,\,window')},\ee the solution to the BFKL
equation  with saturation boundary conditions at $k_{\perp}\sim
Q_s$ is well approximated by
 the scaling form \cite{SCALING,MT02,MP03} :
\be\label{BFKLfix}\varphi_A(\kk,y)\approx
 \frac{1}{\alpha_s N_c} \,\left\{\ln \frac{k_{\perp}^2}{Q_s^2(A,y)}\,
+\,\mathcal{O}\big(1\big)
\right\}\left(\frac{Q_s^2(A,y)}{k_{\perp}^2}\right)^\gamma,\qquad
\gamma\approx 0.63.
\ee
The difference $1-\gamma\approx 0.37$ 
is sometimes referred to as an `anomalous dimension'. Indeed, the fact
that $\gamma$ is strictly smaller than one makes the function in
Eq.~(\ref{BFKLfix}) to show weaker dependencies upon $A$ and $1/\kk^2$
than the leading twist approximation (that one could naively expect
to apply at momenta above $Q_s$).
Below, I shall succinctly refer to the window (\ref{swindow1}) as the `BFKL 
regime'.

The standard leading twist (or DGLAP) approximation is recovered only
at even higher momenta $k_{\perp} \gg Q_s^2(A,y)/Q_s(A)$, 
where the gluon occupation factor is extremely small, and the only
trace of saturation is visible in the fact that it is the {\it original}
saturation momentum at $y=0$, i.e., $Q_s(A)$, which acts as the
infrared cutoff for the transverse phase--space available for
evolution. Namely, because of the collinear singularity of QCD
(see, e.g., Eq.~(\ref{bremss})),
the phase--space for the emission of a single (high--$\kk$)
gluon is ${\rm d}\kk^2/\kk^2$, and therefore the total
phase--space available for the evolution from $Q_s(A)$ up to $\kk$
is $\rho(A,\kk) \equiv \ln \kk^2/Q_s^2(A)$.
Within the present approximations, the solution to the linear equation
which resums such small--$x$ {\it and} collinear gluon
emissions is given by
 \be\label{DLAp0}
\varphi_A(\kk,y)\,\simeq\,\frac{\mu_A}{k_{\perp}^2}\,\,
{\rm exp}\left\{\sqrt{4\bar\alpha_s y \rho(A,\kk)}\right\}\,,
\ee
which is recognized as the evolution of the leading--twist
term in the classical distribution (the
first term in the r.h.s. of Eq.~(\ref{phiA0})). Eq.~(\ref{DLAp0}) is
known as the ``double--logarithmic accuracy'', or DLA, approximation,
and is strictly valid so long as  $\bar\alpha_s y \rho \gg 1$.

By using the previous formulae, together with similar formulae for
the proton, 
it is possible to compute the ratio
$\mathcal{R}_{pA}(\kk,y)$ and follow its evolution with $y$. The
corresponding formulae for the case of a running coupling
$\alpha_s\equiv\alpha_s(\kk^2)$ are quite similar\footnote{More
complete formulae for the case of a running coupling can be found
in the original literature \cite{SCALING,MT02,DT02,MP03}, and are
also summarized in Ref. \cite{IIT04}.} (at least, within the
kinematical ranges in which Eqs.~(\ref{phiAsat}) and
(\ref{BFKLfix}) apply, i.e., at saturation and in the BFKL
regime), with one crucial difference though: the functional form
of the saturation momentum, which for running coupling reads
\cite{SCALING,MT02,AM03} \be\label{Qsrun}
Q_s^2(A,y)\,\simeq\,\Lambda^2_{\rm QCD} \,\,{\rm
exp}\left\{\sqrt{2c b y+\rho_A^2}\right\}\,,\ee
where $b\equiv {12N_c}/{(11N_c -2N_f)}$ and
$c\simeq \,4.88$ is the same number as in Eq.~(\ref{Qsfix0}).
As compared to the corresponding formula for a fixed coupling,
Eq.~(\ref{Qsfix0}), the expression (\ref{Qsrun})  shows
a less rapid increase with $y$ at high energies, and also a
weaker dependence upon the atomic number $A$, which 
becomes less important with increasing $y$ \cite{AM03}:
\be\label{QsrunAp}
\frac{Q_s^2(A,y)}{Q_s^2(p,y)}\,\simeq\,\,\exp\left\{\frac{\rho_A^2-\rho_p^2}
{2\sqrt{2c b y}}\right\}\, \qquad{\rm for} \qquad 2c b y\,\gg\, 
\rho_A^2\,>\,\rho_p^2\,,\ee
and eventually disappears:
\be\label{QsrunY} Q_s^2(A,y)\,\simeq\,\Lambda^2_{\rm QCD}
\,{\rm e}^{\sqrt{2c b y}}\,
 \qquad{\rm for} \qquad 2c b y\,\gg\, \rho_A^4\,.\ee
For sufficiently large energies, quantum evolution with running
coupling washes out completely the difference between a nucleus
and a proton !

\section{V. \
 Forward rapidities: High--$\lowercase{p}_\perp$ suppression}

We are now prepared for a study of the evolution of the ratio
$\mathcal{R}_{pA}(\kk,y)$ with increasing $y$, starting with the
initial condition provided by the MV model (which exhibits Cronin
enhancement at intermediate momenta, as discussed in Sect. III).

\subsection{V.1. \ Quantum evolution of $\mathcal{R}_{pA}$
: General features}

Let me first summarize the main features of the evolution, and then
develop some of them in the next subsections:

a) The main effect of the evolution is a {\it rapid suppression}
of the ratio $\mathcal{R}_{pA}$, due to
the {\it different evolution rates} for the
gluon distributions in the {\it nucleus} (the
numerator in Eq.~(\ref{Rdef})) and in the {\it proton}
(the denominator there). The proton distribution grows faster 
because, for the same values of $\kk$ and $y$, the transverse 
phase--space available 
for its evolution is larger than that for the nucleus. 
Indeed, as noticed in Sect. IV,
the transverse phase--space $\int^{\kk}\big({\rm d}^2p_\perp/
p_\perp^2\big)$ is limited by the infrared cutoff
introduced by the initial conditions at $y=0$. For the nucleus,
this cutoff is the relatively hard scale $Q_s(A)$ associated
with {\it saturation\,}; thus, the nuclear phase--space
$\rho(A,\kk) \equiv \ln \kk^2/Q_s^2(A)$ is considerably smaller
than the proton one, $\rho(p,\kk) \equiv \ln \kk^2/Q_p^2$ with 
$Q_p\sim \Lambda_{\rm QCD}$. Specifically,
 \be\label{rhop}
 \rho(p,\kk)\,\equiv\,\ln \frac{\kk^2}{Q_p^2} \,=\,
\ln\frac{ \kk^2}{Q_s^2(A)} + \ln\frac{Q_s^2(A)}{Q_p^2}\,\simeq\,
\rho(A,\kk) + \rho_A,\ee where $\rho_A \gg 1$ has been generated
according to Eq.~(\ref{rhoA}). Correspondingly, 
${\mathcal R}_{pA}(\kk,y)$ decreases very fast with $y$, and 
already after a short evolution\footnote{Recall that 
$\bar\alpha_s(Q_s^2(A)) = b/\rho_A \sim 1/\rho_A$ since 
$b= \mathcal{O}(1)$.} 
$ y \sim 1/(\bar\alpha_s\rho_A)\, \sim 1$ it becomes
smaller than one at all but the asymptotic momenta.

b) By the same argument, the suppression goes away at extremely
large momenta, where the difference between $Q_s(A)$ and $Q_p$
becomes unimportant in computing the phase--space. In fact, when
$\kk\gg Q_s(A,y)$, one can use the DLA formula (\ref{DLAp0}) for
both the proton and the nucleus, and thus deduce (for fixed
coupling) :
 \be {\mathcal R}_{pA}(\kk,y)\,\simeq \,{\rm
e}^{-\rho_A\sqrt{Y/\rho(A,\kk)}}\qquad{\rm for}\qquad \kk\gg
Q_s(A,y)\,,\ee which approaches one {\it from below} when
$\kk\to\infty$.

c) The {\it suppression rate} $\,{\rm d} \ln{\mathcal R}_{pA}/{\rm
d}y\,$ is largest at small $y\,$ and for not so large transverse
momenta [say, for $\,k_{\perp} \simle Q_s^2(A,y)/Q_s(A)\,$], 
since in this regime the dissymmetry between the
evolution of the proton and that of the nucleus is most
pronounced: The proton is in the DLA regime, and thus evolves very
fast (because of the large transverse phase--space available to
it), whereas the nucleus shows geometric scaling, and
evolves only slowly (because, so long as $\bar\alpha_s y < 1$, the
nuclear saturation momentum rises very little; see
Eq.~(\ref{Qsfix0})).

This explains, in particular, the rapid suppression in ${\mathcal
R}_{pA}$ observed in the early stages of the evolution in the
numerical study in Ref. \cite{Nestor03}.

d) For larger $y$ such that $\bar\alpha_s y\simge 1$, the ratio
${\mathcal R}_{pA}(\kk,y)$ is {\it monotonously increasing} with
$\kk$. That is, the Cronin peak has flattened out during the
first $1/\bar\alpha_s$ units of rapidity.

e) The {\it flattening} of the Cronin peak cannot be attributed to
the proton evolution {alone} --- the latter produces a
quasi--uniform suppression in ${\mathcal R}_{pA}$ at momenta around
$Q_s(A,y)$, so, by itself, it would preserve a local structure like a
peak---, rather this must be related to the evolution of the nucleus. As
we shall see, it is indeed the {\it nuclear evolution} which
washes out that distinguished feature of the initial distribution
which was responsible for the existence of a well--pronounced peak at
$y=0$ : the exponential fall off of the gluon occupation factor at
momenta just above the saturation plateau (cf. Sect. III).

f) Whereas the {\it generic} features of the evolution, as
described above, are qualitatively similar for both fixed and
running coupling, important differences persist between these two
scenarios as far as the {\it details} of the evolution, and also
the precise structure of the final results, are concerned.
Specifically, after including running coupling effects, the
evolution appears to be {\it slower} (one needs a larger increase
in rapidity to achieve a given suppression in $\mathcal{R}_{pA}$),
but eventually {\it stronger} (the final value for
$\mathcal{R}_{pA}$ which is obtained after a very large evolution
in $y$ is significantly smaller with running coupling than with
fixed coupling). Let me be more specific on these two points:

To appreciate how {\it fast} is the evolution, let me introduce
the rapidity $y_0$ after which the ratio ${\mathcal
R}_{pA}(\kk,y)$ at $\kk\sim Q_s(A,y)$ decreases from its initial
value of ${\mathcal O}(\rho_A)$ (cf. Eq.~(\ref{RmaxMV})) to a
value of ${\mathcal O}(1)$. In the next subsection, we shall see
that
 \be\label{Ry0}
 y_0\,\simeq\,\frac{1}{4\bar\alpha_s}\,\frac{\ln^2 \rho_A}{\rho_A}\,\sim\,
\frac{\big(\ln  \ln A^{1/3}\big)^2}{\ln A^{1/3}}\,\qquad{\rm
 (fixed\,\,coupling)}\,,\ee
(which incidentally is a very small rapidity interval:
$\bar\alpha_s y_0\ll 1$), and, respectively,
 \be\label{y0run}
 y_0\,\simeq\,\frac{1}{4b}\,\ln
\rho_A\,\sim\,\ln A^{1/3}\qquad{\rm (running\,\,coupling)},\ee
which for large $A$ is parametrically larger than the fixed
coupling estimate in Eq.~(\ref{Ry0}); thus, the running of the
coupling slows down the evolution.

Furthermore, to characterize the {\it strength} of the suppression
after a very large rapidity evolution, consider the limit of
${\mathcal R}_{pA}$ when $y\to \infty$ with fixed $z\equiv
k_{\perp}^2/Q_s^2(A,y)$. (This is the meaningful way to take the
large--$y$ limit, since the interesting physics is located around
the nuclear saturation momentum.) For $z= {\mathcal O}(1)$, one
finds:
 \be{\mathcal R}_{pA}(z\sim
1,y\to\infty)\,\sim\,
\frac{1}{(A^{1/3}\rho_A)^{1-\gamma}}\,\qquad{\rm
 (fixed\,\,coupling)} \,,
 \label{RpA_DSW_yc} \ee
and, respectively,
 \be\label{Rmaxlimit} {\mathcal R}_{pA}(z\sim
1,y\to\infty)\, =\,
  \frac{1}{A^{1/{3}}} \,\,\qquad{\rm (running\,\,coupling)}
 \,.\ee
As anticipated, for large $A$, the running coupling result
(\ref{Rmaxlimit}) is much smaller than the corresponding one for
fixed coupling, Eq.~(\ref{RpA_DSW_yc}) (recall that $1-\gamma\simeq
0.37$).

In fact, the power of $A^{1/3}$ in the r.h.s. of
Eq.~(\ref{Rmaxlimit}) is simply the factor introduced by hand in
the definition (\ref{Rdef}) of ${\mathcal R}_{pA}$. That is, the
result (\ref{Rmaxlimit}) arises directly from the observation
that, with a running coupling and for sufficiently large $y$, the
nuclear and proton saturation scales coincide with each other, cf.
Eq.~(\ref{QsrunY}), so the corresponding occupation factors will
coincide as well, in the whole kinematic range for geometric
scaling (which includes the saturation domain at $z\le 1$
and the BFKL regime (\ref{swindow1})).

g) The dependence of the ratio ${\mathcal R}_{pA}$ upon $A$ is
also interesting, since this corresponds to the centrality
dependence of the ratio $R_{\rm dAu}$ measured at RHIC
\cite{Brahms-data,QM04}. Consider the $A$--dependence for momenta
around the Cronin peak: Whereas at $y=0$, the ratio ${\mathcal
R}_{pA}(\kk\sim Q_s(A))$ is logarithmically {\it increasing}
with $A$ (recall Eq.~(\ref{RmaxMV})), this tendency is rapidly
reversed by the evolution (see the next subsection) : After only a
small rapidity increase $y \sim 1/(\bar\alpha_s\rho_A)\, \sim 1$, 
${\mathcal R}_{pA}(\kk,y)$ becomes a {\it decreasing} function of $A$ for any
$\kk$, in qualitative agreement with the corresponding change in the
centrality dependence observed in the data, cf. Fig.
\ref{brahmsCP}.

\subsection{V.2. The suppression of the Cronin peak}

I shall now use some simple calculations to illustrate the prominent role
played by the proton evolution for the suppression of
${\mathcal R}_{pA}$ (especially at small $y$). For definiteness,
I shall consider the fixed coupling case and focus on
transverse momenta of the order
of the nuclear saturation momentum, since this is the region where
the Cronin peak is located at $y=0$. (The generalization of the
discussion below to a running coupling and to arbitrary transverse 
momenta can be found in Ref. \cite{IIT04}.) 
So, let's consider the evolution with $y$ of the following quantity
(the ratio (\ref{Rdef}) along the nuclear saturation line) :
\be\label{Rsat} {\mathcal R}_{\rm sat}(A,y)\equiv {\mathcal
R}_{pA}(\kk=Q_s(A,y),y)\,.\ee
Along this line, the nuclear gluon distribution reduces to a constant,
due to geometric scaling (cf. Eq.~(\ref{phiAsat})) : 
\be\label{philine} \varphi_A(\kk=Q_s(A,y),y) \sim 1/{\alpha_s N_c}\,.\ee
As for the corresponding distribution
in the proton, this is given by the linear (BFKL) evolution,
since $\kk=Q_s(A,y)\gg Q_s(p,y)$. Specifically, for relatively
small $y$, such that $\bar\alpha_s y \ll \rho_A$, the evolution 
is dominated by the large transverse logarithm $\rho(p,\kk)\sim 
\rho_A$ (cf. Eq.~(\ref{rhop})), and the proton
is described by the DLA formula (\ref{DLAp0}), 
while for $\bar\alpha_s y\simge  \rho_A$ the proton enters the BFKL regime,
where Eq.~(\ref{BFKLfix}) applies. The `critical' rapidity
$\bar\alpha_s y_c \sim \rho_A$ at which the proton changes from  
DLA to BFKL is the upper limit for the geometric scaling
window (\ref{swindow1}), now applied to the proton.

\bigskip
I) $\bar\alpha_s y \ll \rho_A$ : {\it The early stages of the evolution
(proton at DLA)}
\bigskip

Using Eq.~(\ref{DLAp0}) with $A\to p$ and $\kk=Q_s(A,y)$, together
with $\rho(p,Q_s(A,y))\simeq \rho_A$ 
(cf. Eq.~(\ref{rhop})) and the definition (\ref{rhoA}), 
one immediately finds:
\be\label{RmaxDLAfix} {\mathcal R}_{\rm sat}(A,y)\,
\,\sim\, \rho_A \,\,
{\rm exp}\Big\{c\bar\alpha_s y-\sqrt{4\bar\alpha_s y\rho_A}\,\Big\}.
\ee
For $y=0$, this is parametrically large,
${\mathcal R}_{\rm sat}(A,y=0) \sim\rho_A$,
as expected (cf. Eq.~(\ref{RmaxMV})). But when increasing $y$
(with $\bar\alpha_s y \ll \rho_A$ though), the ratio decreases 
very fast --- the DGLAP increase of the proton
distribution being faster than the BFKL increase of the nuclear
saturation momentum, cf. Eq.~(\ref{Qsfix0}) ---, and becomes
parametrically of $\mathcal{O}(1)$ already after the very short
rapidity evolution
\be\label{Ry01} \bar\alpha_s y_0\,\sim\,\frac{\ln^2
\rho_A}{4\rho_A}\,
\,\ll\,1\,,\ee
which is Eq.~(\ref{Ry0}). This value $y_0$
is so small that one can in fact ignore the corresponding 
evolution of the nucleus:
{\it The rapid decrease in the height of the peak in the
very early stages of the evolution is entirely due to the DGLAP
evolution of the proton.} 

In fact, Eq.~(\ref{RmaxDLAfix}) shows that, larger is
$\rho_A$ (i.e., higher was the original peak at $y=0$), faster
is the suppression seen when increasing $y$ 
(i.e., smaller is $y_0$). This is so since $\rho_A$ also fixes the
transverse phase--space for the DGLAP evolution of the proton, 
and as such it
enters the exponential factor in Eq.~(\ref{RmaxDLAfix}). For the
same reason, the ratio (\ref{RmaxDLAfix}) turns rapidly into
a {\it decreasing} function of $\rho_A$ (and
thus of $A$), as anticipated in Sect. V.1.

\bigskip
I) $\bar\alpha_s y \simge  \rho_A$ : {\it Proton in the
BFKL regime}
\bigskip

For $y>y_c$, the proton enters the scaling window (\ref{swindow1}),
where Eq.~(\ref{BFKLfix}) becomes appropriate. I shall shortly
argue that, when this happens, the Cronin peak has already
disappeared; but it is still interesting to follow the ratio
${\mathcal R}_{pA}(\kk,y)$ further up along the nuclear saturation
line. By using Eq.~(\ref{BFKLfix}) with $Q_s(A,y)\rightarrow
Q_s(p,y)$, together with Eq.~(\ref{philine}) and
the relations (which follow
from Eqs.~(\ref{QsatMV}), (\ref{rhoA}) and $\mu_A = A^{1/3}\mu_p$) 
 \be\label{QSAPY}\frac{Q_s^2(A,y)}{Q_s^2(p,y)}\,= \, \frac{Q_s^2(A)}{Q_p^2}
 \,\simeq \,A^{1/3}\,{\rho_A}\,,\qquad
\ln\frac{Q_s^2(A)}{Q_p^2}\,\simeq\,\rho_A,\ee
one easily finds
 \be\label{RmaxBFKLfix} {\mathcal R}_{\rm
sat}(A,y)\,\sim\,\frac{1}{(A^{1/3}\rho_A)^{1-\gamma}}
\,\,\ll\,1\,,\ee 
as anticipated in Eq.~(\ref{RpA_DSW_yc}).
Note the power of $A$ in the denominator: this provides a strong
suppression factor which is independent of $y$. This power was
missing in DLA (compare to Eq.~(\ref{RmaxDLAfix})), but
appears here as a consequence of the `anomalous dimension' $1-\gamma
> 0$ characteristic of the BFKL solution in the vicinity of the
saturation line \cite{SCALING,MT02}.

Note furthermore that the result in Eq.~(\ref{RmaxBFKLfix}) is
independent of $y$ : the $y$ dependencies have cancelled 
in the ratio (\ref{QSAPY}) between the nuclear and
the proton saturation momenta. Thus, as compared to the (proton) 
DLA regime at $y < y_c$, where ${\mathcal R}_{\rm
sat}(A,y)$ is rapidly decreasing with $y$, in the BFKL regime at
$y > y_c$ this ratio stabilizes at a very small value,
proportional to an inverse power of $A$.

At this level, one can easily anticipate that the behaviour
at large $y$ should be quite different with a running coupling:
Indeed, in that case, the $y$--dependencies do {\it not} 
compensate in the ratio ${Q_s^2(A,y)}/{Q_s^2(p,y)}$ (cf.
Eqs.~(\ref{Qsrun})--(\ref{QsrunAp})), so the 
function ${\mathcal R}_{\rm sat}(A,y)$ keeps decreasing also in the
(proton) BFKL regime, down to the limiting value 
 ${\mathcal R}_{\rm sat} = 1/A^{1/3}$, cf. Eq.~(\ref{Rmaxlimit}).
In practice, this value is reached
for $2c b y\,> \rho_A^4$ (cf. Eq.~(\ref{QsrunY})).

\subsection{V.3. The flattening of the Cronin peak}

To study the fate of the Cronin peak with increasing $y$,
one needs to extend the previous analysis to momenta outside (but
near to) the saturation line $\kk=Q_s(A,y)$. In this region, 
the rapid evolution of the proton provides a strong suppression,
as already discussed, but by itself this evolution is quite uniform 
in $\kk$ and it could preserve a local peak (whose height would be
rapidly decreasing though). From Fig.~\ref{brahmsR}, we note that
the current data are not accurate enough to tell us whether the
peak actually persists with increasing $y$, or not. But from a
theoretical perspective we expect this peak to rapidly flatten out
and disappear after only a short evolution, 
for reasons to be explained shortly \cite{IIT04}. Such a
behaviour has been first seen numerically, in Ref. \cite{Nestor03}, 
where the BK equation \cite{B,K} has been solved with initial 
conditions of the MV type (cf. Sect. III). This scenario
is further supported by the analytic estimates in Sect. IV, 
which show that, when extrapolated down to $\kk\sim Q_s(A,y)$, 
the BFKL solution (\ref{BFKLfix}) becomes {\it parametrically large},
of $\mathcal{O}(1/\bar\alpha_s)$, and thus can be directly matched
onto the corresponding extrapolation of the solution at saturation,
Eq.~(\ref{phiAsat}). This property suggests that there is no
room left for a (parametrically enhanced) peak around $Q_s(A,y)$.

To better appreciate why this matching issue is relevant for the problem
of the flattening of the Cronin peak, one should recall from Sect. III 
that the existence of a well pronounced peak at $y=0$ is precisely 
related to the large {\it mismatch}, around
$\kk= Q_s(A)$, between the `saturating' distribution
$\varphi_A^{\rm sat}$ and the `twist' distribution
$\varphi_A^{\rm twist}$ (the sum of the power--law tails
at high momenta) : For $\kk\sim Q_s(A)$,
$\varphi_A^{\rm sat}\sim 1/\bar\alpha_s$ is 
parametrically larger than $\varphi_A^{\rm twist}
\sim 1/(\bar\alpha_s\rho_A)\sim 1$ (cf. 
Eqs.~(\ref{phiA0}) and (\ref{phiAlow})). Because of this
mismatch, there is a sudden drop (actually, an
exponential fall--off) in the nuclear gluon distribution from
the saturation plateau at low $\kk$ to the power--law tail
at high $\kk$. This rapid fall--off is responsible for the well 
pronounced peak in the ratio $\mathcal{R}_{pA}$
visible in Fig.~\ref{RpAMV}.

One can now understand why the quantum evolution leads to
the flattening of the Cronin peak: Since non--local in $\kk$, 
the evolution replaces any exponential profile in the initial 
conditions (here, the one at momenta just above $Q_s$) 
by a power--law profile. Thus, with increasing $y$, the 
`exponential gap' between the saturation plateau and the power--law 
tail is rapidly filled up, predominantly due to radiation from those 
gluons which were originally at saturation. Correspondingly,
the peak flattens out, and completely disappears after a (rather
short) rapidity evolution $\bar\alpha_s y\sim 1$ \cite{IIT04}.

\begin{figure}[tbp]
\centering{
\includegraphics[scale=1.2]{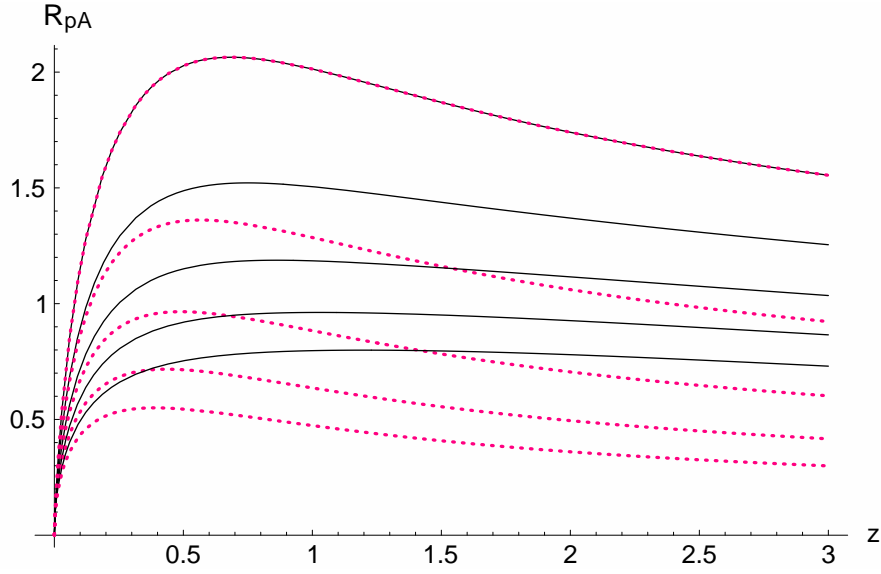}
    \caption{\label{tilt} {\small The Cronin ratio
    $\mathcal{R}_{pA}(z)$ (with $z\equiv \kk^2/Q_s^2(A)$)
 below and near the saturation scale for
    $\rho_A=6$. The solid lines correspond to an
    evolved nuclear wavefunction by $\Delta Y \ll 1$. The
    dotted lines correspond to an unevolved one (MV). The proton
    wavefunction is always given by the DLA solution. The
    curves, from top to bottom, correspond to $\Delta Y=n/(2
    \rho_A)$ with $n=0,1,...,4$.}}
}
\end{figure}

To explicitly follow the flattening of the peak, one cannot rely
on the approximate solutions presented in Sect. IV --- the latter 
apply only for larger $y$ with $\bar\alpha_s y > 1$, and merely 
show that the Cronin peak has already disappeared by then ---, 
but rather one needs more accurate solutions.
In the analysis in Ref. \cite{IIT04}, the BK equation 
has been iterated exactly (analytically) to follow
the first few steps in the evolution of the nucleus. By using
this evolution together with the DLA approximation for the proton 
distribution, one obtains the results displayed in  Fig. \ref{tilt} 
for $\rho_A=6$ (the length of a step in
rapidity is taken to be $\Delta Y=1/(2\rho_A)$
with $Y\equiv \bar\alpha_s y$). For comparison, Fig. \ref{tilt}
also shows the ratio which is obtained
when the non--evolved, MV model, distribution is used for the
nucleus. The rapid suppression of the peak, due to the fast rise
in the proton distribution, is clearly seen in both cases. But in
the absence of nuclear evolution the peak is always there; just
its amplitude gets smaller and smaller. By contrast, when using
the properly evolved nucleus distribution, the
flattening of the peak is manifest, and in fact the maximum has
almost disappeared already after an evolution $\Delta Y=2/\rho_A \approx
0.3$.

\begin{theacknowledgments}

I would like to thank the organizers of the 
International Workshop IX Hadron Physics and VII Relativistic Aspects of
Nuclear Physics (HADRON-RANP 2004, Angra dos Reis, Brasil, 
March 28--April 03, 2004), for their kind invitation and a most exciting
meeting. Also, I wish to express my gratitude to Eduardo Fraga 
and the Theoretical Physics Department of Universidade Federal 
do Rio de Janeiro for hospitality during my visit there, prior
to the Workshop.

\end{theacknowledgments}


\bibliographystyle{aipproc}   


\end{document}